# *Monopole Embedded Eigenstate in Nonlocal ɛ-Near Zero Nanostructures*

**Filipa R. Prudêncio[1,2][*], and Mário G. Silveirinha[1]**

[1]University of Lisbon – Instituto Superior Técnico and Instituto de Telecomunicações, Avenida Rovisco Pais 1, 1049-001 Lisbon, Portugal

[2]Instituto Universitário de Lisboa (ISCTE-IUL), Avenida das Forças Armadas 376, 1600-077 Lisbon, Portugal

## Abstract

In recent years, the confinement of light in open systems with no radiation leakage has raised great interest in the scientific community, both due to its peculiar and intriguing physics and due to its important technological applications. In particular, materials with near-zero permittivity offer a unique opportunity for light localization, as they enable the formation of embedded eigenstates in core-shell systems with suppressed radiation loss. For all the solutions presented thus far in the literature, the exact suppression of the radiation leakage can occur only when the size of the resonator is delicately tuned. Surprisingly, here it is shown that the tuning of the resonator radius may be unnecessary, and that nonlocal metal spherical nanostructures of any size may support multiple embedded eigenstates with a monopole-type symmetry.

---

[*] Corresponding author: filipa.prudencio@lx.it.pt

# Main Text

In the last decade, bound states with suppressed radiation loss have received a great attention by the optics community [1-11]. Surprisingly, it has been shown that carefully designed open material resonators may support localized excitations known as "embedded eigenstates", which in ideal conditions – in the absence of material dissipation– do not decay in time. This result is particularly intriguing considering the fact that the system is open and the eigenstate "lives" in the radiation continuum. Similar "embedded eigenstates" can occur in other types of wave platforms, e.g., in the form of electronic states with "positive energy" in condensed matter systems [12-15].

Conventional optical material structures, e.g., formed by standard dielectrics, can support embedded eigenstates if and only if they are infinitely extended in space [16]. In truncated dielectric systems, the coupling between the trapped light and the radiation continuum is unavoidable, and thereby the lifetime of an excitation is necessarily finite. Remarkably, it was shown in Ref. [16] that materials with a near-zero permittivity offer a unique opportunity to break this very fundamental restriction, and that in theory they enable the formation of ideal "embedded eigenstates" with no radiation loss. Different configurations of this class of embedded eigenstates were studied by several groups [17-24]. Importantly, a recent work demonstrated that the inevitable nonlocality (spatial dispersion) of the ε-near-zero response does not spoil the formation of embedded eigenstates, but, on the contrary, creates additional degrees of freedom and new opportunities for light-trapping [23]. Counterintuitively, it was shown that the nonlocal effects due to the electron-electron repulsive interactions in a metal may considerably relax the conditions for the observation of embedded eigenstates in plasmonic nanostructures [23]. Nonetheless, similar to the local case [16], the formation of the embedded eigenstate is due to a pole-zero cancellation of a Fano-type resonance, and thereby requires the delicate tuning resonator dimensions [23].

The class of ε-near-zero resonators studied previously [17-24] consist of core-shell nanostructures, with the core a dielectric and the shell an ε-near-zero material that effectively shields the light trapped in the core from the outer space (radiation continuum). The trapped energy is stored both at the shell and at the core. The lowest-order embedded eigenstate has a dipolar-type symmetry in the core. Higher-order modes (e.g., quadrupolar, etc) are also allowed, but modes with a global monopole symmetry are strictly forbidden. The shape and size of the resonator depend on the ε-near-zero material and also on the order (often determined by symmetry) of the embedded eigenstate. In particular, for a given resonator shape, the formation of embedded eigenstates requires a delicate tuning of the resonator size, and thus the quality factor of the resonance may be highly sensitive to fabrication imperfections.

Here, we explore a totally different and simpler class of open plasmonic resonators without the inner dielectric core. We show that a spherical resonator made of a material with a nonlocal plasmonic response described by the hydrodynamic model [25] supports multiple embedded eigenstates independent of its size. Thus, different from all the other solutions explored thus far [17-24], the embedded states as robustly protected against perturbations in the resonator size. The embedded eigenstates are longitudinal waves, reminiscent of bulk plasmons. The oscillation frequency of the embedded eigenstates is on the order of the plasma frequency of the bulk material.

Figure 1(a) depicts the geometry of a spherical plasmonic nanoparticle with radius $R$ standing in air. The nanoparticle may be a noble or alkali metal at optical frequencies [26, 27] or a narrow gap semiconductor at terahertz frequencies. We model the wave interactions in the plasmonic material (electron gas) with the hydrodynamic model, which corresponds to the Maxwell's equations coupled to a transport equation and to the charge continuity equation [23, 25-35]:

$$\nabla \times \mathbf{E} = -\mu_0 \partial_t \mathbf{H}, \qquad \nabla \times \mathbf{H} = \mathbf{j} + \varepsilon_0 \partial_t \mathbf{E}, \tag{1a}$$

$$\partial_t \mathbf{j} = \varepsilon_0 \omega_p^2 \mathbf{E} - \omega_c \mathbf{j} - \beta^2 \nabla \rho, \tag{1b}$$

$$\partial_t \rho + \nabla \cdot \mathbf{j} = 0. \tag{1c}$$

Here, $\rho$ and $\mathbf{j}$ are the charge and current density, respectively, associated with the free-electrons, $\omega_p$ is the plasma frequency, $\omega_c$ is the collision frequency, and $\beta$ is a velocity that controls the strength of the diffusion term $-\beta^2 \nabla \rho$. The physical origin of the diffusion term is the repulsive electron-electron interactions in the material that prevent the localization of charge. The value of $\beta$ can be estimated from $\beta^2 = 3/5 v_F^2$, with $v_F$ the Fermi-velocity of the electron gas [37]. Typical values for $\beta$ are on the order of $c/450$ in alkali metals [26] and $c/280$ in semiconductors [36].

The solutions of the drift-diffusion model of the electron gas [Eq. (1)] may be classified as transverse waves ($\nabla \cdot \mathbf{E} = 0$, $\rho = 0$ and $\mathbf{H} \neq 0$) and longitudinal waves ($\nabla \cdot \mathbf{E} \neq 0$, $\rho \neq 0$ and $\mathbf{H} = 0$). Here, we look for longitudinal-type embedded eigenstates in spherical resonators. For $\mathbf{H} = 0$ the curl of the electric field vanishes, and thereby it can be written as a gradient of an electric potential: $\mathbf{E} = -\nabla \phi$. Since the magnetic field vanishes, it is clear from Eq. (1a) that $\mathbf{j} = -\varepsilon_0 \partial_t \mathbf{E}$. Then, considering a time-harmonic variation, $\mathbf{E} = \mathbf{E}_\omega e^{-i\omega t}$, $\mathbf{j} = \mathbf{j}_\omega e^{-i\omega t}$, $\phi = \phi_\omega e^{-i\omega t}$, it is straightforward to check that the electric potential must satisfy:

$$\nabla^2 \phi_\omega + \frac{1}{\beta^2}\left(\omega^2 - \omega_p^2 + i\omega\omega_c\right)\phi_\omega = 0. \tag{2}$$

For a spherical geometry the solutions of the above equation can be constructed using spherical harmonics:

$$\phi_\omega = \phi_0 \, j_n(k_L r) Y_n(\hat{\mathbf{r}}), \qquad k_L = \frac{1}{\beta}\sqrt{\omega^2 - \omega_p^2 + i\omega\omega_c}. \tag{3}$$

Here, $\phi_0$ is an arbitrary normalizing factor, $j_n$ is the spherical Bessel function of order $n$ ($n = 0, 1, ...$) and $Y_n(\hat{\mathbf{r}})$ is a spherical harmonic of order $n$. The electric field can be written as:

$$\mathbf{E}_\omega = -\phi_0 \left[ k_L j'_n(k_L r) Y_n(\hat{\mathbf{r}}) \hat{\mathbf{r}} + j_n(k_L r) \frac{1}{r} \text{Grad} Y_n(\hat{\mathbf{r}}) \right]. \tag{4}$$

The electromagnetic fields in the outer free-space region must vanish for an ideal embedded eigenstate with infinite lifetime [16]. Thus, in order to ensure the continuity of the tangential electromagnetic fields at the interface $r = R$ it is necessary that $\hat{\mathbf{n}} \times \mathbf{E}_\omega = 0$ and $\hat{\mathbf{n}} \times \mathbf{H}_\omega = 0$, with $\hat{\mathbf{n}}$ the unit normal vector. Furthermore, diffusion effects prevent the accumulation of charges at the interface, and as a consequence the normal component of the electric current density in the electron gas must vanish at the interface $\mathbf{j}_\omega \cdot \hat{\mathbf{n}} = 0$ [23, 25-35]. Since $\mathbf{j}_\omega = i\omega \varepsilon_0 \mathbf{E}_\omega$ this additional boundary condition is equivalent to state that $\hat{\mathbf{n}} \cdot \mathbf{E}_\omega = 0$. Thereby, we conclude that the total electric field must vanish at the interface with the free-space region ($\mathbf{E}_\omega = 0$).

From Eq. (4), the constraint $\mathbf{E}_\omega = 0$ can be satisfied for $n \geq 1$ (dipole-type and higher-order spherical harmonics) only if $j_n(k_L r)$ and $j'_n(k_L r)$ vanish simultaneously. It is well-known that $j_n, j'_n$ do not have common zeros different from zero, and thereby there are no solutions with $n \geq 1$. On the other hand, since the zero-order ($n = 0$) spherical harmonic is a constant ($Y_0(\hat{\mathbf{r}}) = const.$), it follows that the modal equation for an embedded eigenstate with monopole-symmetry reduces to $j'_0(k_L r) = 0$, or equivalently $j_1(k_L R) = 0$. The function $j_1(u)$ has an infinite number of positive roots $u = u_m$ ($m = 1, 2, ...$), being the first few roots $u_1 = 4.49$, $u_2 = 7.73$, .... Solving $k_L R = u_m$ with respect to the frequency, one finds that the oscillation frequency of the $m$-th embedded eigenmode is:

$$\omega = \omega_m \equiv \sqrt{\frac{u_m^2}{R^2}\beta^2 + \omega_p^2 - \left(\frac{\omega_c}{2}\right)^2} - i\frac{\omega_c}{2}, \qquad m = 1, 2, \dots . \tag{5}$$

Remarkably, due to the nonlocal effects a spherical nanoparticle supports an infinite number of embedded eigenstates with monopole-symmetry. The electric field profile for the fundamental ($m=1$) state is represented in Fig. 1(b). As seen, the electric field is radial and its intensity is peaked at some finite distance from the center of the nano-sphere.

Importantly, the previous analysis shows that the formation of eigenstates does not require tuning the size of the nano-particle, as the eigenstates can emerge for any radius $R$, rather different from what happens in core-shell resonators [17-24]. All the monopole embedded eigenstates have the same lifetime, $\tau_{\text{life}} = 1/(-2\omega'') = 1/\omega_c$ with $\omega''$ the imaginary part of the oscillation frequency ($\omega = \omega' + i\omega''$). In particular, the lifetime of the embedded eigenstates $\tau_{\text{life}}$ is controlled by the collision (damping) frequency, and is exactly identical to the lifetime of the bulk plasmons in the electron gas. Different from the core-shell resonators considered in previous works, the radiation from the nanoparticle is exactly suppressed even when the system is lossy ($\omega_c \neq 0$), because the monopole mode is inherently protected against radiation loss. In the ideal case where the dissipation vanishes, the oscillations in the nanoparticle do not decay in time and have infinite lifetime.

Figure 2 depicts the oscillation frequency ($\omega = \omega_m$) of the first few branches of embedded eigenstates as a function of the radius of the nanoparticle $R$ (for $\beta/c = 300$) and as a function of the diffusion velocity $\beta$ (for $R\omega_p/c = 1$), in panels a) and b) respectively. In this simulation, the effects of material dissipation are neglected ($\omega_c = 0$). The frequency of oscillation of the embedded monopole eigenmodes varies continuously with the structural parameters of the resonator, guaranteeing that $\omega'' = \text{Im}\{\omega\} = 0$ for any configuration. As seen in Fig. 2a, by adjusting the radius of the nanoparticle it is possible to control the spectral distance between the different embedded eigenmodes, and hence to store light in the resonator at multiple frequencies. A reduction in the size of the resonator causes a blue shift of the oscillation frequency. Likewise, an increase of the diffusion velocity $\beta$ also blue shifts the oscillation frequency. In the local limit, $\beta \to 0$, all the branches collapse into a single branch with $\omega = \sqrt{\omega_p^2 - \left(\frac{\omega_c}{2}\right)^2} - i\frac{\omega_c}{2}$, which is nothing but the oscillation frequency of bulk plasmons in a local electron gas.

From Eq. (4), the electric field associated with the monopole eigenstates is of the form $\mathbf{E}_\omega = E_0 j_1(k_L r)\hat{\mathbf{r}}$. The electric field lines are radial [Fig. 1(b)]. In particular, for the $m$-th embedded eigenstate:

$$\mathbf{E}_\omega^{(m)} = E_0 j_1\left(u_m \frac{r}{R}\right)\hat{\mathbf{r}}. \tag{6}$$

The current density is $\mathbf{j}_\omega^{(m)} = i\omega\varepsilon_0 \mathbf{E}_\omega^{(m)}$, and the charge density is $\rho_\omega^{(m)} = \varepsilon_0 \nabla \cdot \mathbf{E}_\omega^{(m)} = -\varepsilon_0 \nabla^2 \phi_\omega^{(m)} = \varepsilon_0 k_L^2 \phi_\omega^{(m)} \sim j_0(u_m r/R)$.

Figure 3a shows a time snapshot of the radial electric field profile in the spherical nanoresonator for the first few eigenmodes. As expected, the radial field vanishes at $r = R$ to ensure that there is no accumulation of charge at the boundary. The number of extremes and nulls of the electric field increases with the order $m$ of the eigenmodes. Likewise, the volumetric charge density $\rho_\omega^{(m)}$ also has a standing wave type structure, with the number of nulls being determined by the order of the mode (Fig. 3b). This property and the radial nature of the electric field uncover the nature of the monopole eigenstates: they are charge density oscillations inside the spherical nanoparticle, very similar to bulk plasmons. Due to the nonlocal effects, eigenstates associated with a large value of $m$ can have an oscillation frequency $\omega_m' = \text{Re}\{\omega_m\}$ that differs substantially from the plasma frequency $\omega_p$ of the electron gas. As a consequence, the quality factor of the oscillations increases with *m*: $Q^{(m)} = \omega'_m / (-2\omega''_m) = \omega'_m / \omega_c$. Thus, the nonlocal effects enable excitations with enhanced quality factors. Note that $Q^{(m)}/(2\pi)$ gives the number of oscillation cycles during the lifetime of the oscillation [16]. All the modes have the same lifetime ($\tau_{\text{life}} = 1/\omega_c$), but modes with larger *m* have shorter oscillation periods.

It is interesting to analyze the transport of energy in the nanoparticle in a single oscillation cycle. It is simple to check from (1) that in the absence of external sources the energy balance is expressed by the following conservation law (generalized Poynting theorem) $\nabla \cdot \mathbf{S} + \partial_t W = 0$ (for simplicity material dissipation is ignored in the following discussion, $\omega_c = 0$). The Poynting vector and the energy density given by:

$$\mathbf{S} = \mathbf{E} \times \mathbf{H} + \frac{\beta^2}{\varepsilon_0 \omega_p^2} \rho \mathbf{j}, \tag{7a}$$

$$W = \frac{1}{2}\varepsilon_0 \mathbf{E} \cdot \mathbf{E} + \frac{1}{2}\mu_0 \mathbf{H} \cdot \mathbf{H} + \frac{1}{2\varepsilon_0 \omega_p^2} \mathbf{j} \cdot \mathbf{j} + \frac{\beta^2}{2\varepsilon_0 \omega_p^2} \rho^2. \tag{7b}$$

Due to the spatial dispersion effects [38-39], the instantaneous Poynting vector $\mathbf{S}$ and the instantaneous energy density $W$ depend explicitly on the charge and current density of the electron gas. The parcel $\frac{1}{2\varepsilon_0 \omega_p^2}\mathbf{j}\cdot\mathbf{j} + \frac{\beta^2}{2\varepsilon_0 \omega_p^2}\rho^2$ represents the kinetic energy per unit of volume of the electron gas and a potential energy term related to the effects of diffusion. Note that when $\beta \neq 0$ the Poynting vector differs from the standard formula $\mathbf{S}_0 = \mathbf{E} \times \mathbf{H}$, because the diffusion effects enable the transport of energy in a nonradiative way.

For an embedded monopole eigenstate the magnetic field vanishes, and thereby $\mathbf{S} = \frac{\beta^2}{\varepsilon_0 \omega_p^2}\rho \mathbf{j}$ and $W = \frac{1}{2}\varepsilon_0 \mathbf{E} \cdot \mathbf{E} + \frac{1}{2\varepsilon_0 \omega_p^2}\mathbf{j}\cdot\mathbf{j} + \frac{\beta^2}{2\varepsilon_0 \omega_p^2}\rho^2$. As $\rho, \mathbf{j}$ are in quadrature the time-averaged Poynting vector in one cycle vanishes, as it should be. However, the instantaneous

Poynting vector is nonzero because there is a transport of energy associated with the oscillations of the bulk plasmons. Figure 4 shows several time snapshots of the Poynting vector of the fundamental monopole mode ($m=1$), confirming that indeed there is a bidirectional flow of energy inside the nanoparticle, despite the nonradiative character of the oscillation.

The embedded monopole eigenstates cannot be excited with external currents in the free-space region. The reason is that the field created by an arbitrary external current can always be decomposed into transverse electric (TE) and transverse magnetic (TM) radial Mie harmonics associated with spherical harmonics with $n \geq 1$. Thus, the longitudinal monopole mode is totally decoupled from the radiation fields outside the resonator. Due to this reason, the monopole embedded eigenstates cannot be detected as resonances under plane wave illumination. In principle, the monopole mode can be excited by a source placed inside the resonator or alternatively by an energetic electron beam that travels through the material (Cherenkov-type radiation due to a transformation of kinetic energy of the beam into bulk plasmon oscillations) [40].

As previously discussed, a spherical nano-resonator supports multiple embedded monopole eigenstates independent of its size. In principle, this remarkable property is specific of the spherical geometry. In fact, consider an arbitrarily shaped resonator with boundary $\Sigma$ and a possible longitudinal embedded eigenmode with $\mathbf{E} = -\nabla \phi$. The electric potential must satisfy Eq. (2), subject to the boundary condition $\phi_\omega|_\Sigma = const.$ to ensure that $\hat{\mathbf{n}} \times \mathbf{E}_\omega = 0$. In addition, to ensure that there are no currents flowing through the boundary ($\mathbf{j}_\omega \cdot \hat{\mathbf{n}} = 0$) it is necessary that $\partial \phi_\omega / \partial \mathbf{n}|_\Sigma = 0$, i.e., the normal derivative of the potential must vanish at each and every point of the surface. For a generic geometry, the two boundary conditions ($\phi_\omega|_\Sigma = const.$ and $\partial \phi_\omega / \partial \mathbf{n}|_\Sigma = 0$) cannot be satisfied simultaneously: typically only $\phi_\omega|_\Sigma$ or $\partial \phi_\omega / \partial \mathbf{n}|_\Sigma$ can be satisfied independently. In contrast, the spherical symmetry greatly relaxes the constraint $\partial \phi_\omega / \partial \mathbf{n}|_\Sigma = 0$, because for a monopole distribution the condition $\partial \phi_\omega / \partial \mathbf{n}|_\Sigma = 0$ is effectively equivalent to a single scalar constraint $\int \partial \phi_\omega / \partial \mathbf{n}\, ds = 0$, which can be satisfied for some set of discrete frequencies, as shown previously.

In summary, in this Letter we studied a new class of nanoresonators formed by a nonlocal plasmonic material. It was shown that any spherical resonator, independent of its size, supports simultaneously multiple embedded monopole eigenstates associated with charge density oscillations in the nanoparticle. Different from solutions based on core-shell particles, the proposed system does not require the delicate tuning of the resonator radius. Furthermore, the radiation leakage is exactly suppressed even when the material is lossy. In principle, these remarkable properties are specific of the spherical geometry. The lifetime of the embedded eigenstates is controlled by the collision frequency. Due to the monopole nature of the electric field, the embedded eigenstates are totally decoupled from external excitations and can only be pumped either with an internal source or with an electron beam.

**Acknowledgements:** This work is supported in part by the IET under the A F Harvey Engineering Research Prize, and by Fundação para a Ciência e a Tecnologia and Instituto de Telecomunicações under project UID/EEA/50008/2020.

# Figures

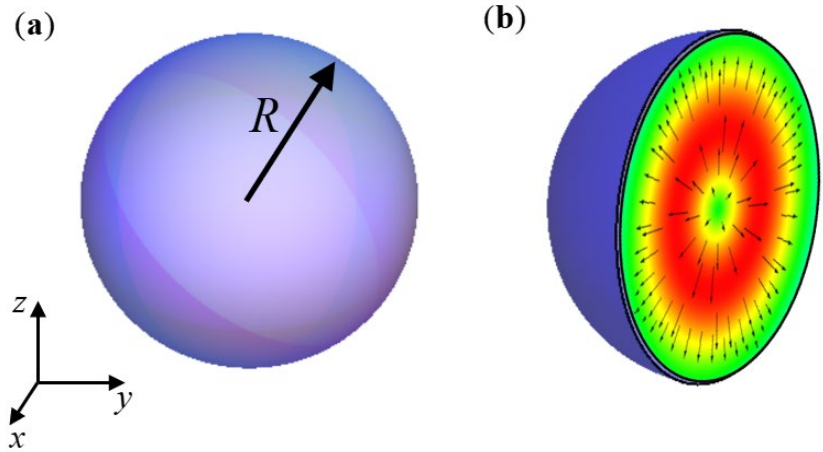

FIG. 1. **(a)** Geometry of a spherical metal nano-particle with a nonlocal response. The radius of the resonator is $R$. **(b)** Time snapshot ($t=0$) of the electric field in the spherical resonator for the embedded eigenstate $m=1$. Red (green) colors represent stronger (weaker) field intensities. Note that the electric field is radial.

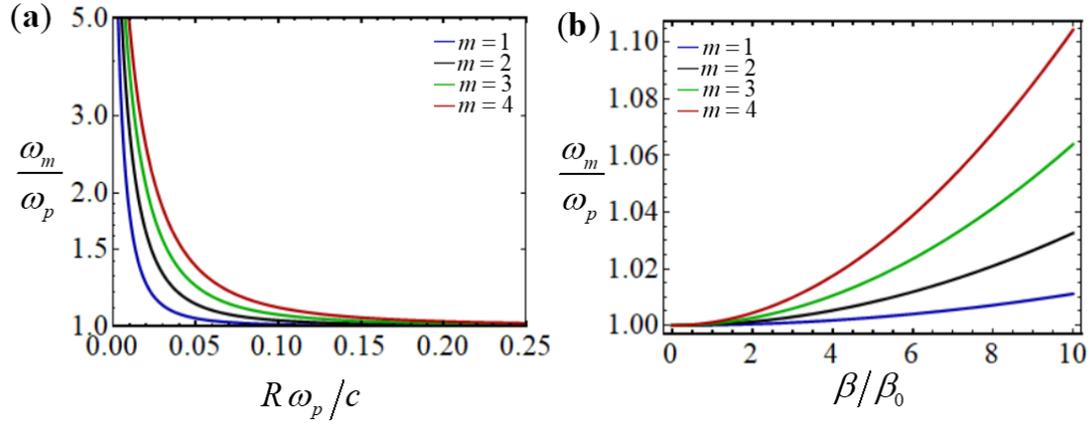

FIG. 2. Real part of the oscillation frequency of the *m*-th embedded monopole eigenstate as a function of **(a)** normalized radius of the nanoparticle $R\omega_p/c$, for a normalized diffusion velocity $\beta/c = 300$. **(b)** normalized diffusion velocity $\beta/\beta_0$ (with $\beta_0 = c/300$), for a normalized radius $R\omega_p/c = 1$. In the plots it is assumed that the collision frequency vanishes $\omega_c = 0$.

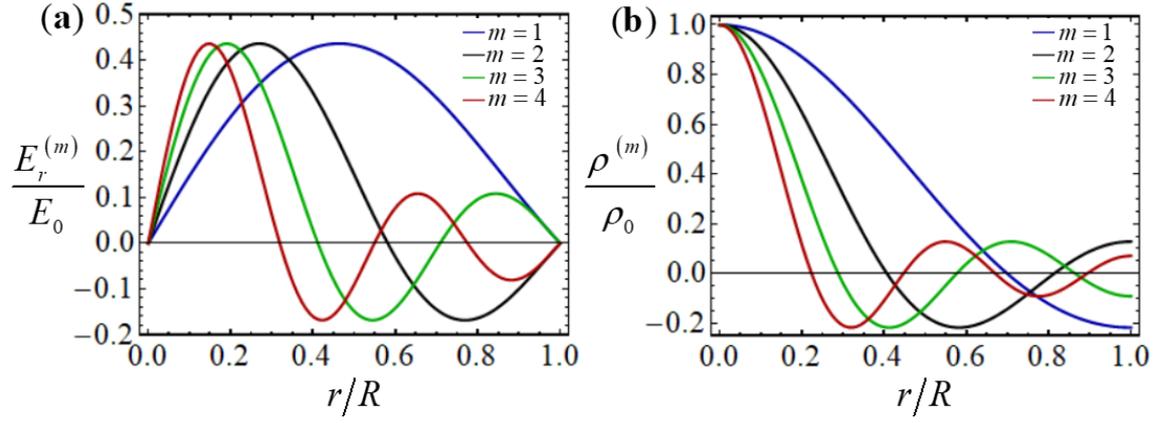

FIG. 3. (a) Snapshot of the normalized electric field in the core ($E_r$) as a function of the normalized radial distance $r/R$ for the $m$-th embedded monopole eigenstate. (b) Snapshot of the normalized charge density in the core ($\rho/\rho_0$) as a function of normalized the radial distance $r/R$ for the $m$-th embedded monopole eigenstate.

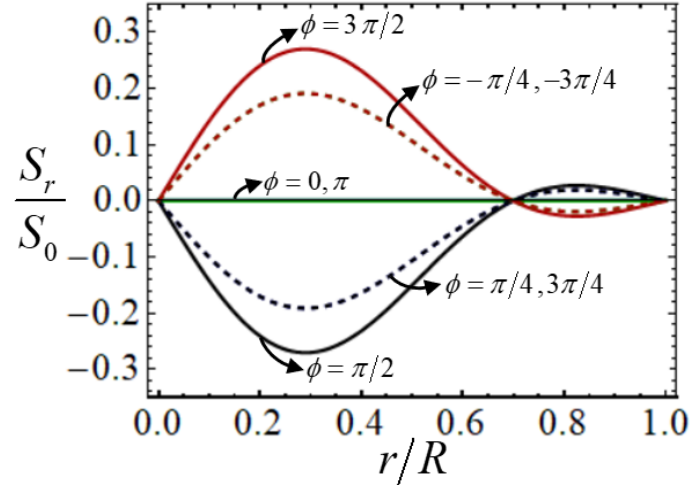

FIG. 4. Time snapshots of the normalized radial Poynting vector component ($S_r/S_0$) for the fundamental embedded monopole eigenstate ($m=1$). The sampling time ($t$) is determined by the value of $\phi = \omega t$. The time-average of the Poynting vector vanishes, but the instantaneous Poynting vector is nontrivial due to the charge density oscillations in the nanoparticle.